\documentclass[twocolumn,aps,prl,preprintnumbers]{revtex4}
\usepackage{bbm}
\usepackage[latin9]{inputenc}
\usepackage{amsmath}
\usepackage{amssymb}
\usepackage{graphicx}
\usepackage{mathrsfs}
\usepackage{amsfonts}
\usepackage{amsthm}
\usepackage{color}
\usepackage{txfonts}
\providecommand{\U}[1]{\protect\rule{.1in}{.1in}}
\makeatletter
\@ifundefined{textcolor}{}
{
\definecolor{BLACK}{gray}{0}
\definecolor{WHITE}{gray}{1}
\definecolor{RED}{rgb}{1,0,0}
\definecolor{GREEN}{rgb}{0,1,0}
\definecolor{BLUE}{rgb}{0,0,1}
\definecolor{CYAN}{cmyk}{1,0,0,0}
\definecolor{MAGENTA}{cmyk}{0,1,0,0}
\definecolor{YELLOW}{cmyk}{0,0,1,0}
}
\makeatother
\begin{document}
\title{Role of atomic spin-mechanical coupling in the problem of magnetic biocompass}
\author{Yunshan Cao}
\email[Corresponding author: ]{yunshan.cao@uestc.edu.cn}
\author{Peng Yan}
\email[Corresponding author: ]{yan@uestc.edu.cn}
\affiliation{School of Electronic Science and Engineering and State Key Laboratory of Electronic Thin Film and Integrated Devices, University of
Electronic Science and Technology of China, Chengdu 610054, China}

\begin{abstract}
It is a well established notion that animals can detect the Earth's magnetic field, while the biophysical origin of such magnetoreception is still elusive. Recently, a magnetic receptor Drosophila CG8198 (MagR) with a rod-like protein complex is reported [Qin \emph{et al}., Nat. Mater. \textbf{15}, 217 (2016)] to act like a compass needle to guide the magnetic orientation of animals. This view, however, is challenged [Meister, Elife \textbf{5}, e17210 (2016)] by arguing that thermal fluctuations beat the Zeeman coupling of the proteins's magnetic moment with the rather weak geomagnetic field ($\sim25-65$ $\mu$T). In this work, we show that the spin-mechanical interaction at the atomic scale gives rise to a high blocking temperature which allows a good alignment of protein's magnetic moment with the Earth's magnetic field at room temperature. Our results provide a promising route to resolve the debate on the thermal behaviors of MagR, and may stimulate a broad interest on spin-mechanical couplings down to atomistic levels.

\end{abstract}
\maketitle
\emph{Introduction.}$-$Magnetoreception is a sense which allows animals, such as salamanders, bees, mole rats, and migratory birds, to detect the Earth's magnetic field to perceive direction, altitude and/or location. How the compass sense works is still a mystery. The solution to this long-standing issue certainly should come from the interplay of physics and biology.

There are two leading physical models to explain the nature of magnetic sensing \cite{Johnsen}. One is the so-called radical pair model which involves the quantum evolution of electron spins with singlet-triplet conversion \cite{Ritz1,Ritz2,Hore,Rodgers,Maeda}: Optical photon excites a spatially separated electron pair in a spin-singlet sate in molecular structures. The anisotropic hyperfine interaction between the electron and the nucleus induces a singlet-triplet conversion. The inclination of the molecule with respect to the geomagnetic field can modulate this conversion and thus can be detected by the radical pair to orient. Cryptochromes are the most promising magnetoreceptor candidates to perceive geomagnetic information via the quantum radical-pair reaction triggered by lights \cite{Rodgers,Maeda}. This mechanism prevents sensing the polarity of the geomagnetic field, but only the inclination. The other model relies on the ferrimagnetism hypothesis \cite{Johnsen}, in which magnetic minerals act as the biocompass to receive and respond to the direction of the Earth's magnetic field. However the identification of such ferrimagnetic organs and/or receptor genes in organisms is difficult \cite{Treiber}. Recently, Qin \emph{et al}. \cite{Xie} discover a magnetoreceptor Drosophila CG8198 (MagR) with a double-helix rod-shaped protein complex with cryptochromes, and co-localizing with cryptochromes, including 40 iron atoms spread out over a length of 24 nm. Their claim subsequently is challenged by Meister \cite{Markus} who argued that the Zeeman interaction between the protein's magnetic moment and the Earth's magnetic field ($\sim25-65$ $\mu$T) is too small (by about $5$ orders of magnitude) to balance the thermal fluctuations at room temperature.

In this work, we initiate a route to clarify the debate by introducing the atomic spin-mechanical coupling \cite{Richardson,Einstein,Barnett,Losby} that is the angular momentum transfer between magnetic and mechanical degrees of freedom, while we are not trying to explain the light activated mechanism \cite{Michael,Clites}. We are motivated by two phenomena observed in Ref. \cite{Xie}: (i) The magnetosensor complex is strongly stretched in case of a good alignment between its long axis and the geomagnetic field, and (ii) the protein crystal would instantly flip 180$^{\text{o}}$ when the polarity of the approaching magnetic field is inverted. They are clear evidences that there exist significant spin-mechanical couplings in the magnetic protein, which unfortunately did not receive sufficient attentions. To highlight the essential physics associated with the spin-mechanical interaction, let's consider a free magnet with magnetic moment $\textbf{M}$ and mechanical angular momentum $\textbf{L}$. In the absence of external magnetic field, the dynamics of $\textbf{M}$ and $\textbf{L}$ must preserve the total angular momentum \cite{Eugene}, i.e., $\textbf{J}=\textbf{L}-\gamma^{-1}\textbf{M}=\text{const.}$, with $\gamma$ the (positive) gyromagnetic ratio. So, a spin-flipping transition $\textbf{M}\rightarrow-\textbf{M}$ is always accompanied by a global rotation of the magnet with a mechanical angular momentum variation $\Delta\textbf{L}=-2\gamma^{-1}\textbf{M}$ and a kinetic energy increase $\Delta E_{k}=(\Delta\textbf{L})^{2}/2I$ (assuming $\textbf{L}=0$ before the spin flipping), where $I$ is the moment of inertia of the magnet. In the case of a cylindrical magnet of radius $r$ and mass $m$ (as schematically shown in Fig. 1), the moment of inertia along the long axis reads $I=mr^{2}/2$. We thus obtain $\Delta E_{k}=(4\gamma^{-2}\textbf{M}^{2}/m)r^{-2}$. This energy must be compensated by the work from external fields, otherwise the spin-flipping process is forbidden by the requirement of energy conservation. So, there exists a blocking temperature
\begin{equation}\label{Blocking}
T_{B}=\frac{4\textbf{M}^{2}}{mk_{B}\gamma^{2}}\frac{1}{r^{2}}
\end{equation}
inversely proportional to the square of the radius of magnetic cylinder under fixed mass and magnetic moment, with $k_{B}$ the Boltzmann constant. For a chain of 40 Fe atoms \cite{Xie}, $|\textbf{M}|\approx219$ $\mu_{B}=2\times10^{-21}$ J T$^{-1}$ \cite{Markus} ($\mu_{B}$ the Bohr magneton), $m\approx3.7\times10^{-24}$ kg (the total mass of 40 iron atoms), and $r\approx r_{\text{Fe}}=0.13$ nm ($r_{\text{Fe}}$ the radius of Fe atom), we estimate the blocking temperature $T_{B}\approx599 \text{ K}$, which is high enough to suppress the thermal fluctuations at room temperature. In the following, we theoretically study
the stochastic dynamics of magnetic moment and rigid-body vectors that are coupled by magnetic anisotropy and Gilbert damping. Our results show that the atomistic spin-mechanical interaction allows a remarkable $30\%$ alignment of the magnetic moment with the geomagnetic field at room temperature. We predict a fast spinning atomic Fe rod/chain inside the magnetic protein. Our results provide a route to resolve the heated debate on MagR.

\begin{figure}[ptbh]
\begin{centering}
\includegraphics[width=0.48\textwidth]{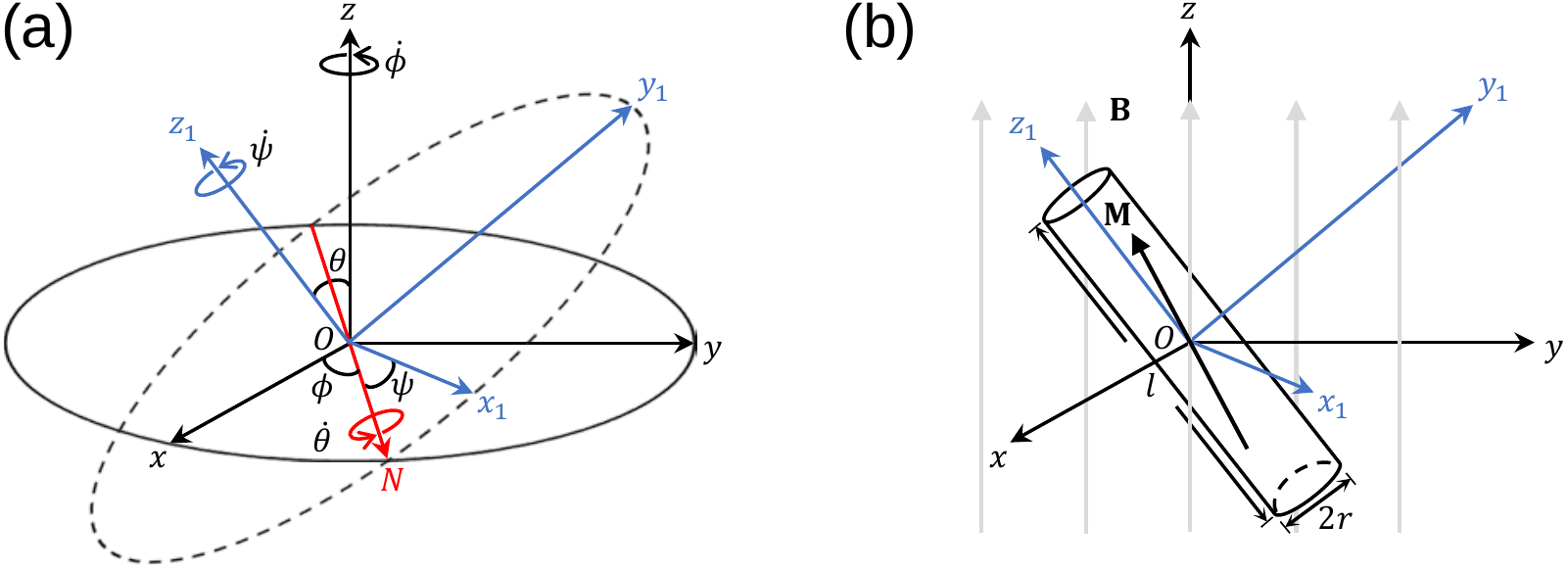}
\par\end{centering}
\caption{(a) Schematic plot of the fixed laboratory frame $(x,y,z)$ and the moving body frame $(x_{1},y_{1},z_{1})$, with origins of the two systems coinciding at point $O$.
The moving $x_{1}y_{1}$-plane intersects the fixed $xy$-plane in some line $ON$ called the line of nodes. Three Eulerian angles $\theta\in [0,\pi],\phi\in [0,2\pi]$, and $\psi\in [0,2\pi]$ measure angles $\angle (\hat{z},\hat{z}_{1})$, $\angle (\hat{x},\hat{N})$, and $\angle (\hat{N},\hat{x}_{1})$, respectively. The angular velocities $\dot{\theta},\dot{\phi}$, and $\dot{\psi}$ are along the $N$-axis, the $z$-axis, and the $z_{1}$-axis, respectively. (b) A cylindrically shaped ferromagnetic needle of length $l$, radius $r$, and magnetic moment $\textbf{M}$ in the (geo-)magnetic field $\textbf{B}\parallel \hat{z}$, with its long axis pointing to the $z_{1}$-direction.}%
\end{figure}
\emph{Model.}$-$We consider a fixed laboratory frame and a moving body frame with their origins coinciding at the $O$ point. The basic vectors of the laboratory frame are $\hat{x},\hat{y},\hat{z}$, and those for the body frame are $\hat{x}_{1},\hat{y}_{1},\hat{z}_{1}$, as shown in Fig. 1(a). Three Euler angles $\theta,\phi$, and $\psi$ measure the rigid-body orientation direction. The body axes are taken to be the principal axes [see Fig. 1(b), in which we approximate the chain of 40 Fe atoms as a rigid magnetic cylinder, for simplicity], along which the tensor of inertia is diagonalized: $I_{1}=I_{2}=m(3r^{2}+l^{2})/12$ and $I_{3}=I=mr^{2}/2$. The long axis is parallel with the $z_{1}$-direction. Since we are only interested in the dynamics at room temperature, a classical description instead of a purely quantum spin model should be well justified: (i) quantum fluctuations from both the Zeeman coupling ($|\textbf{M}||\textbf{B}|/k_{B}\approx4-10$ mK) and the magnetic anisotropy ($D|\textbf{M}|/2k_{B}\approx0.1-1$ K for $D\approx1.5-15$ mT) are far below the room temperature, and (ii) we are treating a spin system with the quantum spin number much larger than $\frac{1}{2}$. Here $D$ is the uniaxial anisotropy constant. The atomistic classical dynamics of magnetic moment $\textbf{M}$ in the laboratory frame is governed by the stochastic Landau-Lifshitz-Gilbert equation \cite{Usov,Hedyeh}
\begin{equation}\label{SLLG}
  \dot{\textbf{M}}=-\gamma\textbf{M}\times \big[\frac{D}{|\textbf{M}|}(\textbf{M}\cdot\hat{z}_{1})\hat{z}_{1}+\textbf{B}_{\text{th}}+\textbf{B}\big]+\frac{\alpha}{|\textbf{M}|}\big[\textbf{M}\times \dot{\textbf{M}}+\textbf{M}\times(\textbf{M}\times\boldsymbol{\Omega})\big],
\end{equation}
where $\hat{z}_{1}=\left(\sin\theta\sin\phi, -\sin\theta\cos\phi, \cos\theta\right)^{\text{T}}$ is the direction of easy axis, $\textbf{B}_{\text{th}}$ is the thermally fluctuating magnetic field with zero average and a time-correlation function satisfying the fluctuation-dissipation theorem FDT \cite{Brown,Yan}:
\begin{equation}\label{FDT}
  \langle B_{\text{th},i}(t)\rangle=0;\langle B_{\text{th},i}(t)B_{\text{th},j}(t')\rangle=\frac{2\alpha k_{B}T}{\gamma |\textbf{M}|}\delta_{ij}\delta(t-t'),
\end{equation}
with $i,j=x,y,z$, $\textbf{B}$ is the weak geomagnetic field along the $z$-direction, $\alpha$ is the phenomenological dimensionless Gilbert damping parameter \cite{Gilbert}, $T$ is the absolute temperature, and $\boldsymbol{\Omega}=(\dot{ \theta}\cos\phi+\dot{\psi}\sin\theta\sin\phi, \dot{ \theta}\sin\phi-\dot{\psi}\sin\theta\cos\phi, \dot{\phi}+\dot{\psi}\cos\theta)^{\text{T}}$ is the angular velocity vector of the rotating cylinder observed in the laboratory frame \cite{Landau}. The mechanical angular momentum in the laboratory frame is $\textbf{L}=R^{\text{T}}\text{diag}\{I_{1},I_{2},I_{3}\}R\boldsymbol{\Omega}$ with a rotational transformation matrix $R$ \cite{Rotation}. The time evolution of the mechanical angular momentum is then determined by
\begin{equation}\label{Lt}
  \dot{\textbf{L}}=\gamma^{-1}\dot{\textbf{M}}+\textbf{M}\times \textbf{B},
\end{equation}
where we have assumed that the Earth's magnetic field $\textbf{B}$ is the only source of angular momentum without considering the mechanical friction. The model (\ref{SLLG}) was originally used to treat the classical magnetic nanoparticles in solution \cite{Usov}, while we adopt the same law of physics to describe the biological system here. Quantum effect may arise in the cases of ultrafast time scales and/or low temperatures, for instance. A rotational wavepacket approach then will be more relevant \cite{Bartels,Ortigoso,Friedrich,Ramakrishna,Jang}.

The set of nonlinear stochastic differential equations (\ref{SLLG})$-$(\ref{Lt}) describe the coupled dynamics of the magnetic moment and the rigid body. According to the FDT (\ref{FDT}), the thermal noise does not play any role in the absence of dissipation ($\alpha=0$), no matter how high the temperature is. The ferromagnetic needle is expected to slowly precess about the geomagnetic field with magnetic moment $\textbf{M}$ being locked with $\textbf{L}$ due to the magnetic anisotropy. In the case of a finite $\alpha$, however, the noise field $\textbf{B}_{\text{th}}$ becomes pronounced at elevated temperatures. It has the tendency to cause a random fluctuation of the magnetic moment $\textbf{M}$ [see the first term in the right hand of Eq. (\ref{SLLG})]. However, we argue that this fluctuation is strongly suppressed in a thin cylindrical magnet ($r\ll l$), as follows: Since $I_{3}\ll I_{1}=I_{2}$, the mechanical rotation around the long axis ($\hat{z}_{1}$) is easiest to be excited, i.e., $\dot{\psi}\gg\dot{ \theta},\dot{\phi}$. So, the angular velocity vector $\boldsymbol{\Omega}\approx \dot{\psi}\hat{z}_{1}$. The strong damping torque [see the second term in the right hand of Eq. (\ref{SLLG})] therefore tends to force the magnetic moment $\textbf{M}$ to be aligned with the long axis $\hat{z}_{1}$. Finally, Eq. (\ref{Lt}) dictates a global precession of the manget body with locked $\textbf{M}$ and $\textbf{L}$ about the Earth's field $\textbf{B}$, immune from thermal fluctuations.

\emph{Numerical method and parameters.}$-$In order to verify our theoretical analysis and to demonstrate the time evolution of the coupled spin-mechanical motion, we solved Eqs. (\ref{SLLG})$-$(\ref{Lt}) numerically at room temperature ($T=300$ K). We set the geomagnetic field strength $|\textbf{B}|=55$ $\mu$T, since the experiments in Ref. \cite{Xie} were performed in Beijing \cite{Beijing}. The corresponding Larmor precession period is $t_{p}=1/(\gamma|\textbf{B}|)\approx$ 0.1 $\mu$s. We adopt the Stratonovich interpretation of the stochastic equation \cite{Stratonovich,Palacios}. Because of the large scale-difference existing between the spin and the rigid-body subsystems, one should do proper parameter rescalings before implementing the numerical simulation. We rescale the time $\mathfrak{t}=t/t_{p}$, so that $\frac{d}{dt}=t^{-1}_{p}\frac{d}{d\mathfrak{t}},\frac{d^{2}}{dt^{2}}=t^{-2}_{p}\frac{d^{2}}{d\mathfrak{t}^{2}},$ and $\delta(t-t')=t_{p}\delta(\mathfrak{t}-\mathfrak{t}')$. The noise is invariable within the $n$th integration step $\Delta \mathfrak{t}$ and is equal to $\textbf{B}_{\text{th}}(\mathfrak{t}_{n})=\gamma^{-1}\sqrt{2\alpha k_{B}T/(|\textbf{M}||\textbf{B}|\Delta \mathfrak{t})}\boldsymbol{\xi}_{n}$ where $\boldsymbol{\xi}_{n}$ is the $n$th realization of a three-component vector with each one being a normal distribution with a unit dispersion. In the simulation, we choose a fixed step $\Delta \mathfrak{t}=5\times10^{-8}$ which corresponds to a real time step $\Delta t=5\times10^{-15}$ s, and consider the initial condition $\theta=6^{\text{o}}$ and $\hat{z}_{1}\parallel\textbf{M}$ at $t=0$. Following rigid-body parameters are adopted: $m=3.7\times10^{-24}$ kg, $r=r_{\text{Fe}}=0.13$ nm, and $l=24$ nm, if not stated otherwise. We set the Gilbert damping constant $\alpha=0.1$. Since the magnetic anisotropy in MagR is unknown, we use $D=15$ mT, a uniaxial magnetic anisotropy in quasi-one-dimensional Fe chains on Pb/Si reported in Ref. \cite{Sun}.

\begin{figure}[ptbh]
\begin{centering}
\includegraphics[width=0.45\textwidth]{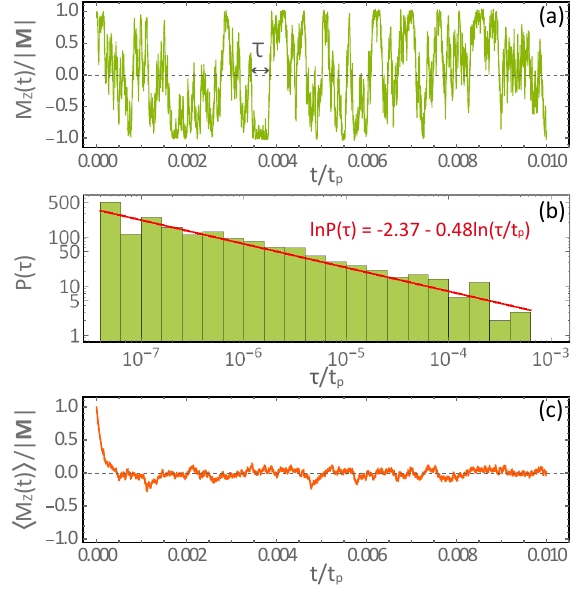}
\par\end{centering}
\caption{(a) Time evolution of $M_{z}$ when the rigid-body's rotational degree of freedom is frozen. Every lifetime $\tau$ between two consecutive switching events has been determined (see the arrow). (b) Respective histogram $P(\tau)$ of the lifetime $\tau$ with $\tau_{\text{min}}=5\times10^{-8}t_{p}$ and $\tau_{\text{max}}=6\times10^{-4}t_{p}$. Red line represents the fitting with a power law. (c) Ensemble average of $M_{z}$ by repeating the simulation $100$ times.}
\end{figure}

\emph{Frozen rigid-body.}$-$We first consider a simple situation that the rotational degree of freedom in the rigid body is frozen, i.e., $\boldsymbol{\Omega}\equiv0$. This corresponds to the case that the iron cylinder is either infinitely heavy, i.e., $I\rightarrow\infty$, or pinned by the protein complex. Figure 2(a) shows the time evolution of the $z$-component of the magnetic moment, from which we find that the thermal fluctuation irregularly flips $M_{z}$. The flipping time is defined as the time between consecutive switching events, with the mean value $\bar{\tau}=\int^{\tau_{\text{max}}}_{\tau_{\text{min}}}P(\tau)\tau d\tau/\int^{\tau_{\text{max}}}_{\tau_{\text{min}}}P(\tau)d\tau$ where $P(\tau)$ is the distribution of $\tau$, and $\tau_{\text{min(max)}}$ is the minimum (maximum) flipping time. Plotting the histogram of all the lifetime $\tau$ as shown in Fig. 2(b), reveals that the lifetime distribution
can be described by a power law $\text{ln}P(\tau)=-2.37-0.48\text{ln}(\tau/t_{p})$, leading to the average lifetime $\bar{\tau}=21$ ps, where we have recorded $1736$ switching events to suppress the statistical error. The obtained mean lifetime can be well understood in terms of the Arrhenius-N\'{e}el-Brown (ANB) formula \cite{Arrhenius,Neel,Brown}
\begin{equation}\label{Arrhenius}
  \bar{\tau}=\nu^{-1}_{0}\text{exp}\big(\frac{E_{b}}{k_{B}T}\big),
\end{equation}
where $\nu_{0}$ is the attempt frequency and $E_{b}$ is the activation energy barrier. \begin{figure}[ptbh]
\begin{centering}
\includegraphics[width=0.45\textwidth]{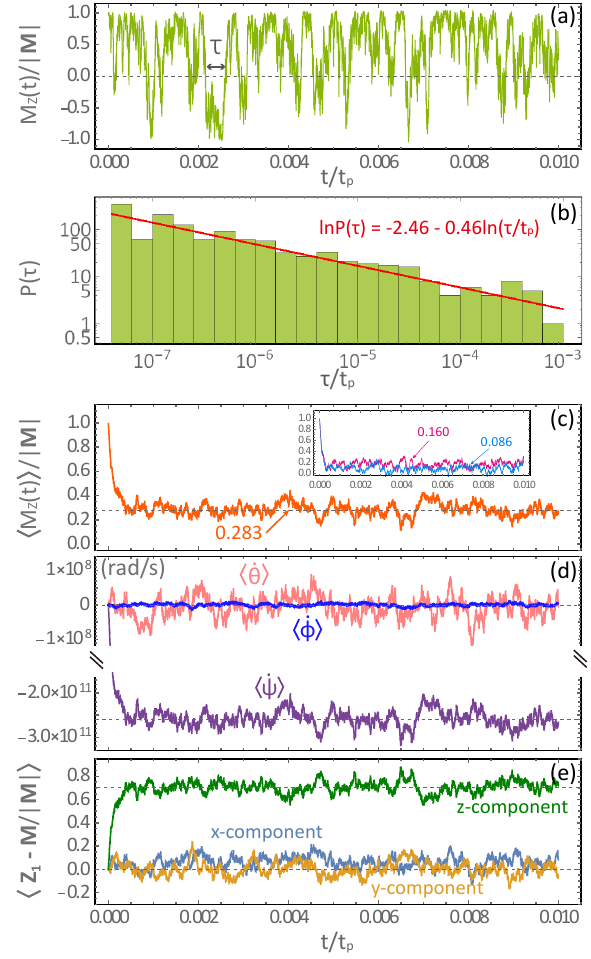}
\par\end{centering}
\caption{(a) Time evolution of $M_{z}$ in the presence of the rotational degree of freedom of the rigid-body. The arrow represents the lifetime $\tau$ between two consecutive switching events. (b) Respective histogram of lifetimes and a power law fitting (red line), with $\tau_{\text{min}}=5\times10^{-8}t_{p}$ and $\tau_{\text{max}}=8\times10^{-4}t_{p}$. (c) Ensemble average of $M_{z}$. (d) Ensemble-average angular velocities of three Euler angles. (e) Time dependence of $\langle\hat{z}_{1}-\textbf{M}/|\textbf{M}|\rangle$. Inset: $\langle M_{z}\rangle/|\textbf{M}|$ for two larger cylinder radii $r=\sqrt{2}r_{\text{Fe}}$ (red curve) and $2r_{\text{Fe}}$ (blue curve).  All ensemble-average quantities are obtained by repeating the simulation for $100$ runs.}
\end{figure}The original estimation of N\'{e}el was $\nu_{0}\approx10^{9}$ Hz, while it is more customary recently to take $\nu_{0}\approx10^{10}-10^{12}$ Hz \cite{Scholz}. Under frozen rigid-body, the activation barrier consists of the anisotropy energy and the Zeeman energy, i.e., $E_{b}=|\textbf{M}|(D+2|\textbf{B}|)/2\approx1$ K multiplying the Boltzmann constant $k_{B}$. We thus get the attempt frequency $\nu_{0}=4.8\times10^{10}$ Hz. The ensemble average of magnetic moment $\langle M_{z}\rangle$ taken over $100$ simulation runs is plotted in Fig. 2(c). It confirms that thermal noises at room temperature indeed beat other interactions \cite{Markus} and completely randomize the magnetic moment, i.e., $\langle M_{z}\rangle\approx0$, in a picosecond time-scale.

\emph{A $30\%$ alignment due to spin-mechanical coupling.}$-$In the following, we investigate the case in the presence of rigid-body degree of freedom. The time evolution of $M_{z}$ is shown in Fig. 3(a), in which irregular spin-flipping phenomena appeared in a similar way to Fig. 2(a). We obtain a longer mean lifetime $\bar{\tau}'=28$ ps by plotting the histogram of lifetimes with the power-law fitting [shown in Fig. 3(b)], where $1215$ switching events have been recorded. The ensemble average over $100$ simulation runs is shown in Fig. 3(c). We find that a remarkable magnetization plateau $\langle M_{z}\rangle=0.283|\textbf{M}|$ emerges [see the dashed line in Fig. 3(c)], after a quick relaxation within tens of picoseconds. Due to the limitation of computing capacities, we only run the simulations for $1$ ns, but we expect that the novel magnetization plateau can last for any longer time since the mechanical friction has been ignored. Because the relaxation process finished in a time scale much shorter than the Larmor period, we deduce that the total angular momentum $\textbf{J}=\textbf{L}-\gamma^{-1}\textbf{M}$ can be viewed as (approximately) conserved [according to Eq. (\ref{Lt})], and thus infer a rigid-body spinning around its long-axis with angular velocity associated with the magnetic moment reduction
\begin{equation}\label{RotationV}
  \dot{\psi}=\frac{\Delta M_{z}}{\gamma I_{3}}\approx-2.5\times10^{11} \text{ rad/s}.
\end{equation}
Calculation of the self-rotation velocity $\langle\dot{\psi}\rangle$ shown in Fig. 3(d) agrees excellently with the theoretical prediction (\ref{RotationV}). Numerical results of the precession frequency $\langle\dot{\phi}\rangle/2\pi\approx-1.8$ MHz as well as the nutation frequency $\langle\dot{\theta}\rangle/2\pi\approx0.02$ MHz are consistent with our previous analysis that they are much smaller than the self-spinning frequency. However, we notice two discrepancies that the calculated average precession frequency $\langle\dot{\phi}\rangle/2\pi$ did not exactly fit the Larmor frequency $1/t_{p}=10$ MHz, and the numerically obtained average nutation frequency $\langle\dot{\theta}\rangle/2\pi$ was not strictly equal to zero. These discrepancies can be resolved by the frequency resolution: The smallest frequency one can resolve in our simulation is $1$ GHz (the reciprocal of the total simulation time $1$ ns). One needs to run the numerical calculation at least $100$ times longer to resolve the geomagnetic Larmor frequency and infinitely long to resolve the (almost) zero nutation frequency, which is not practical and not a center issue for our analysis either. Because the rigid-body's self-rotation acts as an extra energy barrier $\Delta E=I_{3}\dot{\psi}^{2}/2$, the modified ANB law leads to a lifetime
\begin{equation}\label{ArrheniusR}
  \bar{\tau}'=\nu^{-1}_{0}\text{exp}\big(\frac{E_{b}+\Delta E}{k_{B}T}\big)=27 \text{ ps},
\end{equation}
which agrees with the numerical result ($28$ ps) very well. Figure 3(e) shows the relative motion between the rigid-body's long axis and the magnetic moment by plotting the time-dependence of $\langle\hat{z}_{1}-\textbf{M}/|\textbf{M}|\rangle$, from which we find that they are nicely locked with each other [this can be judged by the three (almost) constant projections onto the basic axes of the laboratory frame]. These numerical results confirm our theoretical prediction that atomic spin-mechanical interactions can aid the magnetic moment to resist the thermal fluctuations, allowing a remarkable $30\%$ alignment of the magnetic moment with the rather weak geomagnetic field at room temperature. In our model, the small cylinder radius plays the key role. Structure disorders, however, would cause a larger ``effective" radius, which may somewhat break the alignment. The inset of Fig. 3 indeed shows that the net magnetization has been reduced to $0.16|\textbf{M}|$ and $0.086|\textbf{M}|$ for $r=\sqrt{2}r_{\text{Fe}}$ and $2r_{\text{Fe}}$, respectively.

\emph{Discussion.}$-$We modelled irons in the double-helix rod-shaped MagR protein complex as a free rigid cylindrical magnet and assumed that they are along the rod axis. To construct a more realistic theoretical model, further experimental studies are needed to identify the position of Fe atoms in this magnetosensor polymer, by neutron scattering, for instance. The magnetic moment of the protein is treated as a macrospin in this work. Knowledge about the spin-spin interaction is demanded to improve current theory to explain the origin of the ferromagnetism and to study the biophysics of internal magnetic excitations. The Gilbert damping constant can be measured by magnetic resonance techniques. Its microscopic origin in organisms needs further theoretical investigations. The viscous mechanical damping was neglected in this work, while it can be included into the model accompanied by additional random toques acting on the magnet body, due to the fluctuation-dissipation theorem. Crystalline magnetic anisotropy generally comes from the spin-orbit coupling. Its magnitude in the quasi one-dimensional Fe chains can be determined by means of electron spin resonance.

\emph{Conclusion.}$-$To conclude, we theoretically address the role of the atomistic spin-mechanical interaction in a magnetic chain consisting of tens of Fe atoms, and discover a nice alignment of magnetic moments with the very weak geomagnetic field at room temperature. Numerical results well support our analysis. An important theoretical prediction is the very existence of a self-rotating/spinning atomic Fe rod/chain with angular velocity $\sim10^{11}$ rad/s inside the MagR. Its experimental verification is quite challenging but not completely impossible. One can recur to, for example, rotational Doppler effect techniques \cite{Iwo,Omer} with atomic resolution, to serve that purpose. The relaxation of the atomic rotation is also an interesting open problem. Our findings provide a route toward a resolution of the debate on the thermal properties of MagR and may generate a wide interest on the spin-mechanical interaction at atomic scales.

\begin{acknowledgments}
This work is supported by the National
Natural Science Foundation of China (Grants No. 11704060 and 11604041), the
National Key Research Development Program under Contract No. 2016YFA0300801,
and the National Thousand-Young-Talent Program of China.
\end{acknowledgments}

\end{document}